\documentclass[12pt,article,nofootinbib]{revtex4}
\usepackage{pdfpages}
\usepackage{epsf}
\usepackage{subfigure}
\usepackage{amsmath}
\usepackage{eqnarray}
\usepackage{epstopdf}
\usepackage{mathrsfs}
\usepackage{natbib}
\usepackage{amssymb}
\usepackage{dcolumn}
\usepackage{graphicx}
\usepackage{color} 
\usepackage{hyperref}
\usepackage{enumerate}

\makeatletter
\newcommand*{\rom}[1]{\expandafter\@slowromancap\romannumeral #1@}
\makeatother
\def\be{\begin{equation}}
\def\ee{\end{equation}}
\def\ba{\begin{eqnarray}}
\def\ea{\end{eqnarray}}

\graphicspath{{images/}}
\begin{document}

	\title{\large \bf  Machine learning model to project the impact of Ukraine crisis }

	\author{Javad T. Firouzjaee}
	\affiliation{Department of Physics, K. N. Toosi University of Technology, P. O. Box 15875-4416, Tehran, Iran}
	\affiliation{ School of Physics, Institute for Research in Fundamental Sciences (IPM), P. O. Box 19395-5531, Tehran, Iran } 
		\email{firouzjaee@kntu.ac.ir} 
	
	\author{Pouriya Khaliliyan}
	\affiliation{Department of Physics, K. N. Toosi University of Technology, P. O. Box 15875-4416, Tehran, Iran}
	\affiliation{Qorpi AI Workgroup, P. O. Box 15875-4416, Tehran, Iran}
	\email{pouriya@kntu.ac.ir}

\begin{abstract}
Russia's attack on Ukraine on Thursday 24 February 2022 hitched financial markets and the increased geopolitical crisis. In this paper, we select some main economic indexes, such as Gold, Oil (WTI), NDAQ, and known currency which are involved in this crisis and try to find the quantitative effect of this war on them. To quantify the war effect, we use the correlation feature and the relationships between these economic indices, create datasets, and compare the results of forecasts with real data. To study war effects, we use Machine Learning Linear Regression. We carry on empirical experiments and perform on these economic indices datasets to evaluate and predict this war tolls and its effects on main economics indexes.
 
\end{abstract}
\maketitle

\newpage
\section{Introduction}
It is known that Russia's economy is a mixed economy, with big natural resources, especially oil and natural gas. Russia had 11th rank in nominal GDP in 2021 and is the fifth-largest economy in Europe.
The key feature of Russia is vast geography which is important in its economic activity, and some sources estimate that the nation contains over $ 30\% $ of the world's natural resources \cite{ru-natural}.
Having huge energy resources, Russia has the largest natural gas reserves in the world, the second rank on largest coal reserves, the eighth,  oil reserves. Russia is the main natural gas exporter, especially to Europe regions, the second-largest natural gas producer after the United States.\\

\begin{figure}[htbp!]
	\centering    \includegraphics[width=0.7 \columnwidth]{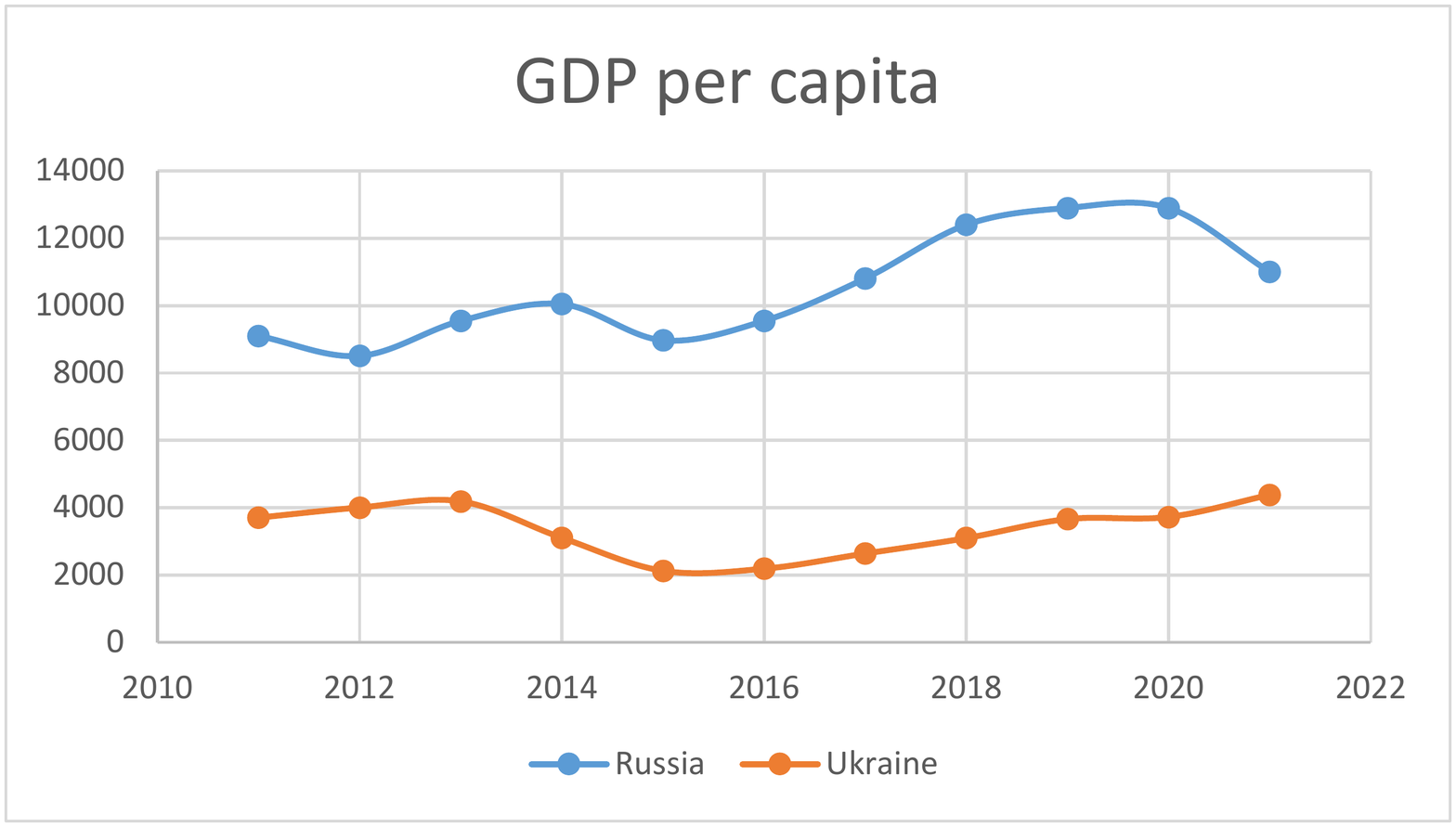}
	\caption{In this picture, we compare the GDP per capita for Russia and Ukraine in the last decade. }
	\label{gdp}
\end{figure}

In contrast, Ukraine had number 56 in the ranking of GDP of the 196 countries in 2020. Ukraine has experienced critical political, security, and economic situations since 2013. The World Bank reported that Ukraine's economic growth rate was $ 2.3\% $ in 2016, thus this was ending the recession \cite{uk-gdp}.  In spite of this progress, the International Monetary Fund reported in 2018, that Ukraine had the lowest GDP per capita of all the countries in Europe. In Fig. \eqref{gdp} you can compare the GDP per capita for Russia and Ukraine in the last decade.\\
To study the Ukraine war economic side-effects, exploring some main economic data indexes and analyzing them can lead us to have better quantitative interpretation.  
Nowadays, it became common to use the Machine learning and Deep Learning approach to analyze economic events. 
Machine learning refers to the study of algorithms and statistical models which subdivision of artificial intelligence that gives computers the ability to learn from data and perform tasks without being explicitly given step-by-step instructions, only having the data it uses to train. Applying different approaches in artificial intelligence,  there are different ways to approach stock price prediction; some of them include technical Analysis, fundamental Analysis, time series analysis. For time series analysis, the Recurrent Neural Network known as one of the effective ways refers to attack the sequential problem or temporal aspects of data as time series which is a powerful tool in stock price prediction \cite{Cen,Pouria,Gupta}. Here, our approach is to use the linear regression method in Machine Learning to study the economic effect of the Ukraine crisis.

The plan of the paper is as follows: Section II is dedicated to an analysis of the correlation between different economic indexes due to the Ukraine war. In Section III, we apply the machine learning approach to predict some main economic indexes involved in this war.  The paper ends in Section VI with a brief summary.\\

\section{Involved economic indexes and their correlation}

\begin{figure}[htbp!]
	\centering    \includegraphics[width=0.7 \columnwidth]{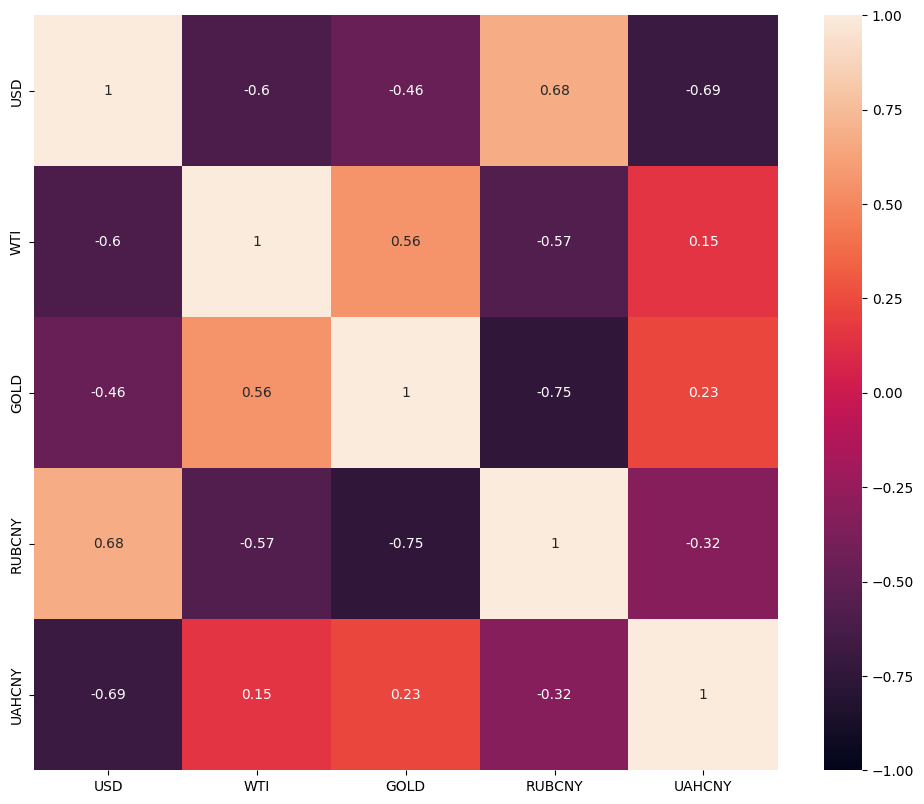}
	\caption{Correlation between USD, WTI, Gold, RUBCNY, and UAHCNY before start Russia and Ukraine crisis (from 2022/01/01 to 2022/02/01). }
	\label{bwar}
\end{figure}

\begin{figure}[htbp!]
	\centering    \includegraphics[width=0.7 \columnwidth]{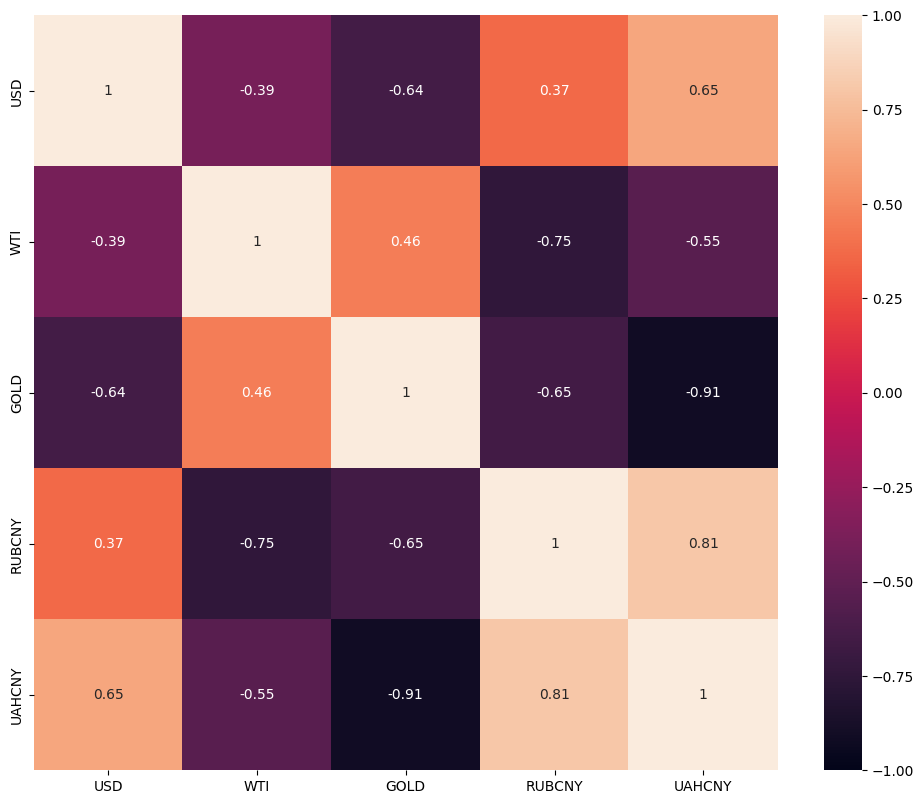}
	\caption{Correlation between USD, WTI, Gold, RUBCNY, and UAHCNY after start Russia and Ukraine crisis (from 2022/02/01 to 2022/03/01). }
	\label{awar}
\end{figure}

In statistics, the dependence of any statistical causal or non-causal ratio between two accidents or data is described by the correlation index. 
This correlation coefficient value is located between -1 and 1 
which negative value or inverse correlation (anti-correlation) expresses a relationship that one value increases as the other decreases. The correlation coefficient is equal to zero if there is no relationship between the two features. In this paper, we select some main economics indexes, such as Gold, Oil (WTI), RUBCNY (RUB/CNY or RUBCNY is the abbreviation for the Russia Ruble and Chinese Yuan pair. It shows how much the RUB (base currency) is worth as measured against the CNY), UAHCNY, and the US dollar index to find out the relationship between features before the Ukraine crisis and after it. Fig. \eqref{bwar} and Fig. \eqref{awar} use Alpha Vantage API data to analyze these economic indexes.

\subsection{Data and Features}

We use following features to predict each stock indexes as following table:
\begin{table}[h]
	\caption{Experiment Result For Royal Dutch Shell Shares}
	\label{rds-table}
	\begin{tabular} {l|l|l|l|l|l|l|l|l}
		& feature1 &feature2 &feature3 &feature4 &feature5 &feature6 &feature7 &Predict feature \\ \hline
		US Dollar index&Open&	High&Low&RUBCNY&UAHCNY&WTI &Gold&Close\\
		NDAQ&Open&	High&Low&Gold&US Dollar &WTI&-&Close\\
		WTI&Open&	High&Low&RUBCNY&UAHCNY&US Dollar &Gold&Close\\
		Gold &Open&	High&Low&RUBCNY&UAHCNY&WTI &US Dollar&Close\\
	\end{tabular}
\end{table}

Our \textbf{training} data is from 2019/05/21 to  2021/06/14, and data from 2021/06/15 to 2021/12/31 is used for \textbf{testing} which extracted from Alpha Vantage API.\\
We use data from 2022/01/03 to 2022/01/31 to predict economics indexes from 2022/02/01 to 2022-03-01 in the case of no war crisis.

\section{War effect estimation from machine learning approach}
\subsection{Machine Learning}
Machine learning is one of the subdivisions of artificial intelligence, in which it examines how the machine learning process is based on mathematical indicators and is known as the correction of implicit calculations. Tom Mitchel defined machine learning \cite{michel}:\\
\textbf{"A computer program is said to learn from experience E with respect to some class of tasks T and performance measure P, if its performance at tasks in T, as measured by P, improves with experience E."}\\

\subsection{Supervised Learning}

Machine Learning can be classified to Supervised learning, Semi-supervised learning, Unsupervised learning, and Reinforcement learning.
In supervised learning, we give the machine the correct answer for each input, to learn and model based on input and output data which enable us to make our own predictions.\\

In data science, regression is a statistical process for estimating the relationships between features. This method includes numerous techniques for modeling and analyzing specific variables, which focus on the relationship between the dependent variable and one or more independent variables. Regression is commonly used for prediction. Furthermore, people use Regression analysis to identify the relationship between the independent and dependent variables and the shape of these relationships. 
In certain circumstances, this analysis can be used to infer excellent relationships between independent and dependent variables.\\

\subsection{Linear Regression}
\begin{figure}[htbp!]
	\centering    \includegraphics[width=0.9 \columnwidth]{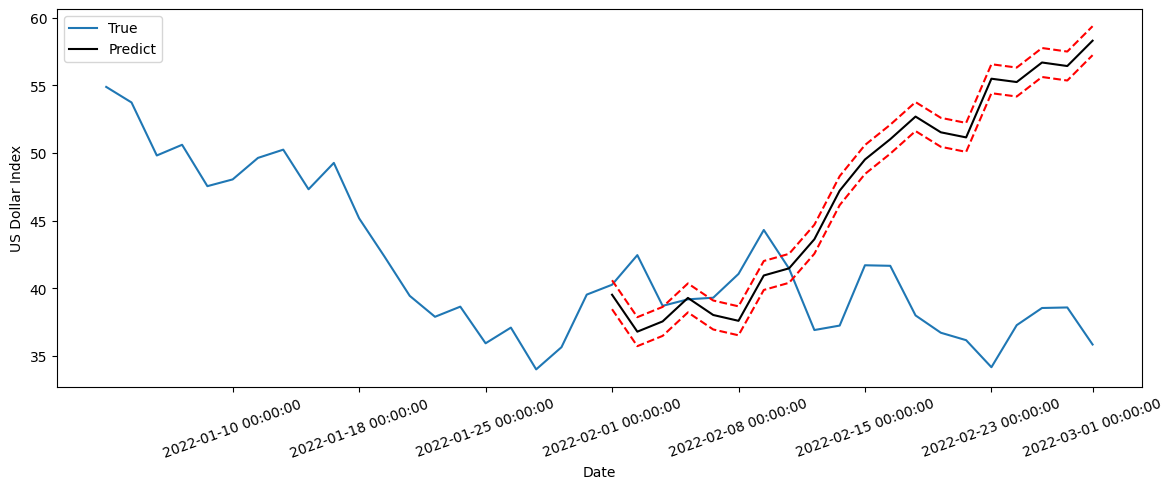}
	\caption{In this graph, we predict the US dollar index in the case of no crisis and compare it with real happened data. }
	\label{usd}
\end{figure}

\begin{figure}[htbp!]
	\centering    \includegraphics[width=0.9 \columnwidth]{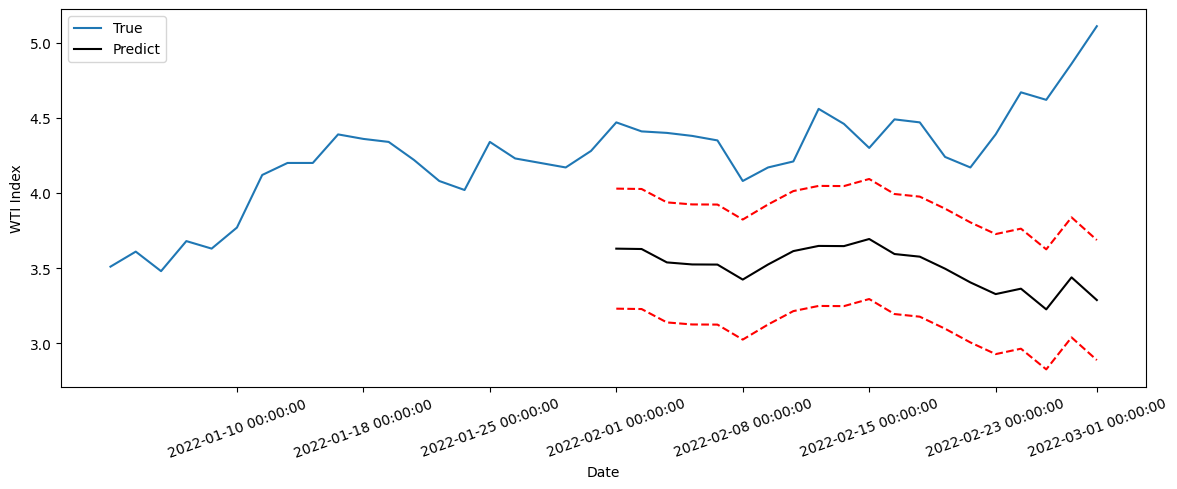}
	\caption{In this graph, we predict the WTI index in the case of no crisis and compare it with real happened data. }
	\label{wti}
\end{figure}

\begin{figure}[htbp!]
	\centering    \includegraphics[width=0.9 \columnwidth]{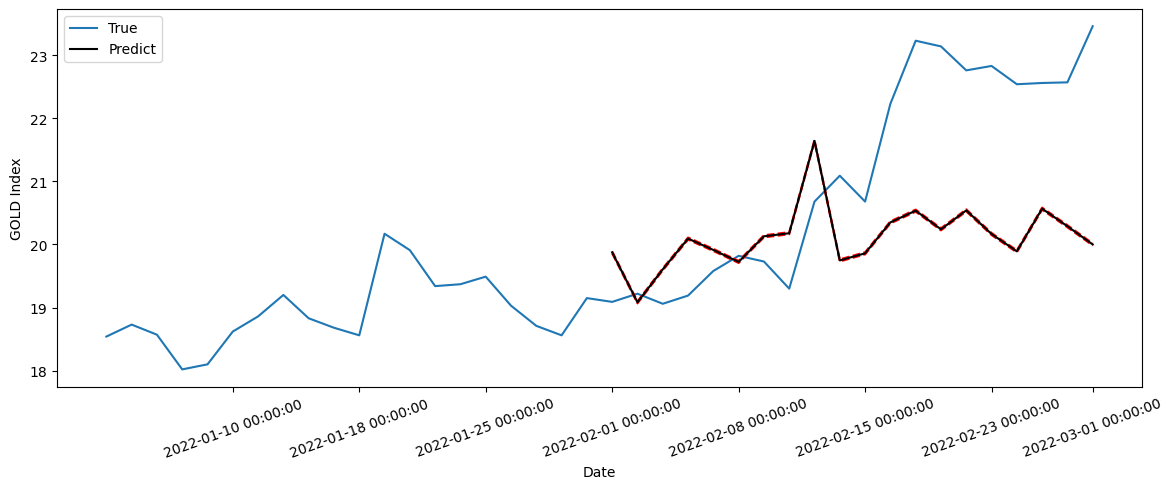}
	\caption{In this graph, we predict the Gold index in the case of no crisis and compare it with real happened data. }
	\label{gold}
\end{figure}

Linear regression is a kind of supervised learning algorithm which measures the effect of an independent variable on a dependent variable and examines the correlations between them.\\
The regression equation which is used to construct a probabilistic model hypothesis can be defined using coefficients $w$ as a linear coefficient and $x$ as an independent variable as follows:\\
\be
h_i=\sum_{0}^{N} w_i x_i+\epsilon,
\ee
where the $\epsilon$ is the error value. The predicted values for the US dollar, WTI, and Gold indexes were depicted in Fig. \eqref{usd}, Fig. \eqref{wti} and Fig. \eqref{gold} in the case of no Ukraine conflict (We designed the machine experience based on task which there is no war).  \\

\subsection{NDAQ}
\begin{figure}[htbp!]
	\centering    \includegraphics[width=0.9 \columnwidth]{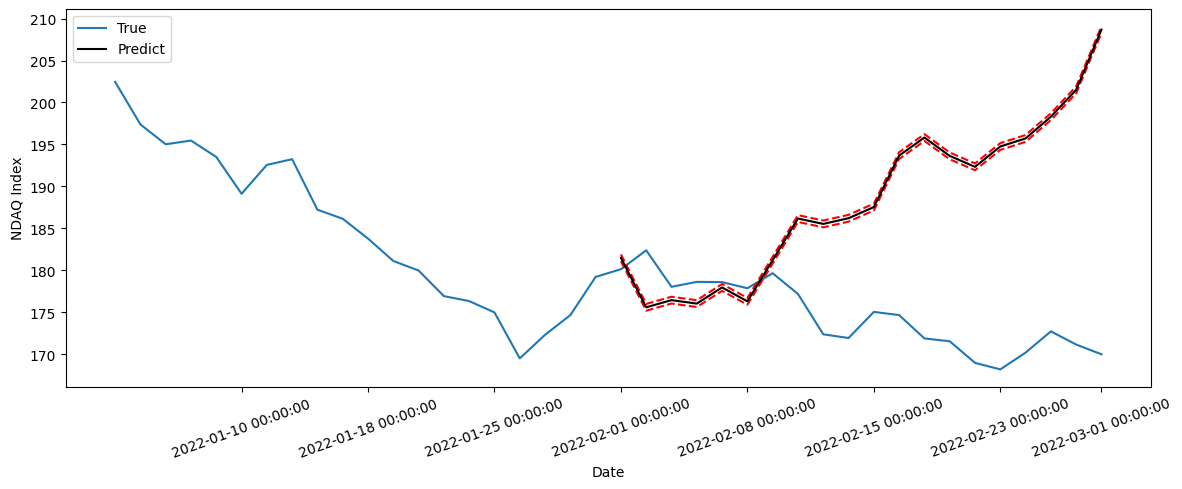}
	\caption{In this graph, we predict the NDAQ index in the case that no crisis and compare it with real happened data. }
	\label{nsaq}
\end{figure}

NDAQ is an American financial holding company that owns a stock market network and several US-based stock and options exchanges and operates three stock exchanges in the US. NDAQ  is headquartered in New York. In  Fig. \eqref{nsaq}, we predict the NDAQ index in the case that no crisis and compare it with real happened data.

\subsection{Evaluation Measures}

There is some evaluation measures that people quantify the loss of the model by Mean Square Error (MSE), Root means square error (RMSE), Mean Absolute Error (MAE), and  Mean Absolute Percentage Error (MAPE). MSE is defined as below:

$$MSE = {{\sum\nolimits_1^n {{{(true - Prediction)}^2}} } \over n}.$$
\\
The average root of the squares of the errors quantified by  RMSE which is defined as:
$$RMSE = \sqrt {{{\sum\nolimits_1^n {{{(true - Prediction)}^2}} } \over n}}. $$
\\
We measure the average of absolute error and use MAE to calculate errors between pair observations which is defined as:
$$MAE = {{\sum\nolimits_1^n {|true - Prediction|} } \over n}.$$
\\
In last, to measure the size of the error in percentage terms people use MAPE which is calculated for the mean of the absolute percentage errors of prediction. It is defined as the following formula:
\be
MAPE = {{{{\sum\nolimits_1^n {|true - Prediction|} } \over {true}}} \over n} \times 100.
\ee
Now, one can apply these evaluation measures as error metrics in our prediction as bellow table.\\

\begin{table}[h]
	\caption{Our Machine Learning model error measurement. }
	\label{rds-table}
	\begin{tabular} {l|l|l|l|l}
		& MSE&	RMSE&	MAE&	MAPE  \\ \hline
		US Dollar index&	1.0711&	1.03494&0.04631&	0.000723\\
		NDAQ&0.40524&	0.03114	&0.00023	&0.63658\\
		WTI&0.39943&	0.632008&	0.48482&	 0.1454\\
		Gold &0.016860&	0.12984&	0.007183&	0.00033\\
	\end{tabular}
\end{table}

\section{Discussion and Conclusion}

Although Russia and Ukraine are responsible for up to $ 30\% $ of the world’s wheat exports and $ 80\% $ of sunflower seed production in the world, according to Capital Economics, this crisis can spark higher global inflation but it is not probable to cause a recession. 
Russia has more than doubled its interest rate to $ 20\% $ in a bid to stop a collapse in the value of its currency which had about $ 30\% $ after one week from the war started.
Although our prediction showed that we could have the US dollar index growth in no conflict case but analysts estimated each $ \$20 $ per barrel increase in oil (one week after starting Russia's attack) would raise core inflation excluding food and energy by 0.035 percentage points and headline inflation by 0.2 percentage points but exacts just a 0.1 percentage point hit to United States GDP \cite{cnbc}. Our correlation analysis shows an increase in the US dollar with RUBCNY  which shows non-significant US involvement in this crisis.

European markets break down at the open amid fears over financial stability, but later decreased losses with London's FTSE 100 closing down $ 0.4\% $, Paris fell $ 1.3\% $, and Frankfurt $ 0.7\% $ \cite{bbc}.

In contrast, analysts say that China could have profit from the diversion of Russian exports if sanctions are imposed on Moscow in the long-term \cite{Aljazeera}. Since during the Ukraine crisis there was no significant change in Chinese currency, we took it as an independent index and normalize the Russia and Ukraine currency to it as RUBCNY and AUHCNY. We can summarise some results as bellow items:

\begin{itemize}
	\item Before the crisis there was a weak correlation between UAHCNY and the gold index, but after the war, this correlation changed to anti-correlation and its value increased.
	
	\item Similar to the Gold-UAHCNY case, before the crisis there was a weak correlation between UAHCNY and WTI index, but after the war, this correlation changed to anti-correlation and its value increased.
	
	\item Before the crisis there was a significant anti-correlation between RUBCNY and the gold index, but after the war, this value did not have a significant change.
	
	\item Before the crisis there was a significant anti-correlation between RUBCNY and WTI index, but after the war, this value becomes greater.
	
	\item Our Machine Learning model predicted that in the case of no crisis,  US dollar index and NDAQ will increase and WTI and Gold will get a small decrease. Whereas, the Ukraine crisis causes that the WTI and Gold indexes increase and  the US dollar index and NDAQ decrease in their index price.
	
\end{itemize}




\end{document}